\documentstyle[epsfig]{l-aa}
\newcommand{\BF}{}
\begin{document} 
\thesaurus{03             
	   (02.13.1;      
	    02.18.5;      
            11.03.4 Coma; 
	    11.09.3;      
	    13.18.2)}     

\title{Limits on Magnetic Fields and Relativistic Electrons in the
Coma Cluster from Multifrequency Observations} 
\author{Torsten A. En{\ss}lin\inst{1,2}
Peter L. Biermann\inst{1,3}}
\institute{Max-Planck-Institut f\"{u}r Radioastronomie, Auf dem 
H\"{u}gel 69, D-53121 Bonn, Germany \and ensslin@mpifr-bonn.mpg.de
\and plbiermann@mpifr-bonn.mpg.de}

\date{Received ??? , Accepted ???}  
\maketitle\markboth{En{\ss}lin \& Biermann: Limits on Magnetic Fields
and Relativistic Electrons in the Coma Cluster}{En{\ss}lin \&
Biermann: Limits on Magnetic Fields and Relativistic Electrons in the
Coma Cluster} 
\begin{abstract}
Relativistic electrons, seen in the large diffuse radio halo of the
Coma cluster of galaxies, should scatter background photons to higher
energies. We calculate the inverse Compton contributions from the
microwave background, from the local radiation field of elliptical
galaxies and from the thermal X-ray emission of the intra cluster
medium to the different observed energy bands, and draw restrictions
to the shape of the spectrum of the relativistic electron
population. The expected electron spectra in different halo formation
models are discussed, and signatures for a future distinction of these
models are presented, tracing the injection processes and the
influence of optical thickness of the radio halo for very low
frequency emission due to synchrotron self absorption. Some of the
upper limits on the electron spectrum translate into lower limits of
the halo magnetic fields, since the observed synchrotron radiation is
known.  Despite uncertainties in the extrapolation of the radio
spectrum to lower frequencies a lower limit to the central field
strength of $B_{\rm o} > 0.3 \,\mu {\rm G}$ seems to be sure. This is 
comparable to the value resulting from minimal energy arguments. If
$B_{\rm o} \le 1.2 \,\mu$G, the recently reported extreme ultraviolet
excess of Coma (Lieu et al. 1996) could result from inverse Compton
scattered microwave photons. Formulae for transrelativistic
Thomson scattering are given in the Appendix.

\end{abstract}
\section{Introduction}
An estimate of a lower limit on the magnetic field strength in the
Coma halo region by Rephaeli et al. (1994)\nocite{rephaeli94} is $B > 0.1
\,\mu {\rm G}$, using the radio spectra of the halo and an upper limit to
inverse Compton (IC) photons in the hard X-ray region measured by the
OSSE experiment. The limit on the field strength derived from the
upper limit to the gamma-ray flux above $100$ MeV by EGRET resulting
from relativistic bremsstrahlung is $B>0.4\,\mu$G (Sreekumar et
al. 1996)\nocite{sreekumar96}, if the radio spectrum can be
extrapolated to lower frequencies.  The field strength, derived by Kim
et al. (1986)\nocite{kim86}, resulting from minimal energy arguments
is $B \approx 0.6 \,\mu {\rm G}\,(1+k_{\rm p})^{2/7}\,h_{50}^{2/7}$,
where $k_{\rm p}$ is the proton to electron energy ratio. Measurements
of Faraday rotation of polarized radiation seen through the Coma
cluster medium, combined with gas profiles, derived from X-ray
observations of the hot intra cluster medium (ICM), give magnetic
fields of $1.7 \,\mu {\rm G}\,(l_{\rm frs}/10\, {\rm
kpc})^{-1/2}\,h_{50}^{1/2}$ (Kim et al. 1990\nocite{kim90}) and
$6.0\,\mu {\rm G} \,(l_{\rm frs}/1\, {\rm kpc})^{-1/2}h_{50}^{1/2}$
(Feretti et al. 1995\nocite{feretti95}).  $l_{\rm frs}$ denotes the
field reversal scale of the magnetic field and can in principle be
measured by depolarization observations. In both measurements the
resolution of $l_{\rm frs}$ was limited by the resolution of the
telescope, and therefore higher field values could result from a
smaller field reversal scale.  Strong magnetic fields on the order of
a few tens of $\mu$G are expected from the injection of radio plasma
from radio galaxies into the ICM (En{\ss}lin et al. 1997).  Although
field limits derived by IC flux limits are lower than the rotation
measurements, they have the advantage of being less independent of the
model. Quantities in this article are scaled to a Hubble constant of
$H_{\rm o} = 50 {\rm\, km \,s^{-1}\, Mpc^{-1}}\,h_{50}$, where
$h_{50}$ indicates their scaling.

\section{Multifrequency Observations}
\subsection{Radio\label{Sec:Radio}}
The radio flux spectrum of the diffuse radio halo Coma C
\begin{equation}
\label{Radio}
F_{\nu} = (8.3 \pm 1.5) \cdot 10^{-12} 
\left( \frac{\nu}{\rm Hz}\right)^{-1.34 \pm 0.06} 
\frac{\rm erg}{\rm cm^{2}\,s\,Hz}
\end{equation}
(Kim et. al 1990\nocite{kim90}) results from a relativistic electron
population, which we assume to be
\begin{equation}
\label{eq:pow}
N_{\rm e} (r, \beta_{\rm e}\gamma_{\rm e})\, d(\beta_{\rm
e}\gamma_{\rm e})  = C_{\rm e}(r)\, (\beta_{\rm e} \gamma_{\rm
e})^{-\alpha}\, d(\beta_{\rm e} \gamma_{\rm e})
\end{equation}
with a spectral index $\alpha = 3.68 \pm 0.12$. $P_{\rm e}= \beta_{\rm e}
\gamma_{\rm e} \,m_{\rm e} \,c$ is the electron momentum, and $\beta_{\rm
e}$ should not be confused with the cluster shape parameter $\beta$.
The synchrotron flux of a power law electron distribution in an
isotropic distribution of magnetic fields within the halo volume
(Eq. (6.36) in Rybicki
\& Lightman 1979\nocite{rybicki79}), averaged over an isotropic 
distribution of electron pitch angles, is
\begin{eqnarray}
\label{Sync}
F_{\nu} = \frac{\sqrt{3} e^3 B_{\rm o} C_{\rm eo}}{m_{\rm e} c^2}
\, f_{\rm sync}(\alpha) \left( \frac{\nu}{\nu_{\rm o}}
\right)^{-\frac{\alpha -1}{2}} \frac{\tilde{V}_{\rm sync}}{4 \pi D^2}\\	
f_{\rm sync}(\alpha) =
 \frac{\sqrt{\pi}\, \Gamma(\frac{\alpha}{4} + \frac{19}{12})\,
\Gamma(\frac{\alpha}{4} - \frac{1}{12}) \, \Gamma(\frac{\alpha+5}{4})}{
2\, (\alpha +1)\, \Gamma(\frac{\alpha+7}{4})}\,\,.
\end{eqnarray}
$B_{\rm o}$, $C_{\rm eo}$ are taken at the cluster center, $D = 139\,
{\rm Mpc}\, h_{50}^{-1}$ is the distance to Coma, $\Gamma$ is the
gamma function, and $\nu_{\rm o} = 3 eB_{\rm o}/(2 \pi m_{\rm e}
c)$. We assume that magnetic field energy density and relativistic
electron density scale radially with the same $\beta$-profile as the
background gas does, $\sim (1+(r/r_{\rm c})^2)^{-3\beta/2}$ with
$\beta= 0.75$ and $r_{\rm c}= 400$ kpc (Briel et
al. 1992\nocite{briel92}). This is reasonable since every possible
source of magnetic fields such as injection by radio galaxies, compression
of primordial fields frozen into the plasma, and amplification of
fields by turbulent gas motion should result in a rough scaling of the
magnetic energy density with the thermal energy density. Similar
arguments hold for the supply of the relativistic electron
population. But the best support of this assumption is given by Fig. 3
of Deiss et al. (1997), which compares the radial X-ray profile with
the radio profile of the cluster, and shows that the latter is
steeper. The X-ray emission scales with the gas density as $\sim
n_{\rm e}^2(r)$, and the radio emission as $\sim C_{\rm
e}(r)\,B^{(\alpha+1)/2}(r) \sim n_{\rm e}^{2.17}(r)$, and is therefore
a little bit steeper, if we use the scaling assumed above ($B^2(r)
\sim C_{\rm e}(r) \sim n_{\rm e}(r)$). The resulting projected emission
profile is similar to that measured by Deiss et al. (1997). The
emission profile itself is shown in Fig. \ref{fig:OptDept}.

The emission weighted volume within the halo radius, which is at least
$R = 1.2\,{\rm Mpc}\,h_{50}^{-1}$ (Deiss et al. 1997)\nocite{deiss97},
is given by
\begin{equation}
\tilde{V}_{\rm sync} = 2 \,\pi\, \phi_{B}\, r_{\rm c}^3\,
{\cal B}_{\scriptscriptstyle \frac{R^2}{R^2+r_{\rm c}^2}}
{\textstyle \left( \frac{3}{2}, \frac{3}{2}
\left( \frac{\alpha+5}{4} \beta -1 \right) \right)}\,\, , 
\end{equation}
where ${\cal B}_x(a,b)$ denotes the unnormalized incomplete beta
function (Eq. (6.6.1) Abramowitz \& Stegun 1965\nocite{abramowitz65}),
and should not be confused with the symbols for the field strength
$B(r)$ and $B_{\rm o}$. $\phi_{B}$ is the filling factor of the
magnetic field in the volume occupied by the relativistic
electrons, which we expect to be close to unity. Comparing Eq.
\ref{Radio} and Eq. \ref{Sync} gives
\begin{equation}
\label{eq:Ceo}
C_{\rm eo} = 3.28 \cdot 10^{-3}\,{\rm cm^{-3}}\,h_{50}\, \phi_B^{-1}\,
 \left(\frac{B_{\rm o}}{\mu{\rm G}}
\right)^{-2.34}\,\,.
\end{equation}

\subsection{Infrared}
Wise et al. (1993)\nocite{wise93} searched extensively for diffuse
far-infrared emission from clusters, and detected Coma marginally at
60 $\mu$m, but not at 100 $\mu$m. Their estimate of the excess fluxes
above the background within 30 arcmin of the cluster center are $5\pm
4$ and $1\pm 10 \,{\rm mJy\,arcmin^{-2}}$. (1 arcmin = 39 $
h_{50}^{-1}$ kpc). We use two sigma upper limits in the following,
namely 13 and 21 ${\rm mJy\,arcmin^{-2}}$.

\subsection{Extreme Ultraviolet\label{Sec:EUV}}
Recently, Lieu et al. (1996)\nocite{lieu96} have reported the
detection of extended, extreme ultraviolet emission from the cluster
center with the Extreme Ultraviolet Explorer (EUVE). They interpret
this as evidence for relatively cold gas components with temperatures
of $kT = 0.07$ and 0.4 keV. But Dixon et al. (1996)\nocite{dixon96}
tried to detect resonance line emission with the Hopkins Ultraviolet
Telescope (HUT), which would be expected in this case. Their
nondetection does not reject the thermal gas model completely, but
gives additional constraint on the properties of the cooler ICM
components.  The EUV emission might also arise from inverse Compton
scattering of microwave or starlight photons, giving then direct
informations about the relativistic electron population. On the other
hand, if the real nature of this emission can be proven to be
different from IC, then we can interpret their flux as an upper limit to
any possible IC contribution, and therefore giving an upper limit to
the electrons. If the detailed physics of such an emission process
can be understood, this limit could be improved.

Their EUVE observation covered a field with a radius of 30 arcmin
centered on the X-ray center of Coma, and they detected emission
within the central 15 arcmin. The count rate in the passband of 65 to
245 eV was 36 \% above the expected count rate from the emission of
the thermal gas with kT = 8.2 keV. We estimate the latter flux in this
band to be $8.1 \cdot 10^{-2}
\,{\rm cm^{-2}\, s^{-1}}$, using their emission measure, which is in
good agreement with that expected from the $\beta$-model of the gas
with a central gas density of $3 \cdot 10^{-3}\,{\rm
cm^{-3}}\,h_{50}^{-1/2}$, and using the Bremsstrahlung formulae given
in Novikov \& Thorne (1972)\nocite{novikov72}. This translates to an
excess flux of $2.9
\cdot 10^{-2} \,{\rm cm^{-2}\, s^{-1}}$, which we use as an upper
limit to the IC flux in the 65 to 245 eV passband in the following.

\subsection{High Energy X-Rays and Gamma-Rays}
Rephaeli et al. (1994)\nocite{rephaeli94} measured the high energy
X-ray (HEX) flux from Coma with OSSE to be lower than $6 \cdot 10^{-6} {\rm
cm^{-2}\,s^{-1}\,keV^{-1}}$ within the 40-80 keV band.  The EGRET flux
limit on gamma rays from Coma is $F_{\gamma}(E_\gamma >\, 100 \,{\rm
MeV}) < 4\cdot 10^{-8} {\rm cm^{-2}\, s^{-1}}$ (Sreekumar et al.
1996\nocite{sreekumar96}).

\begin{figure*}
\psfig{figure=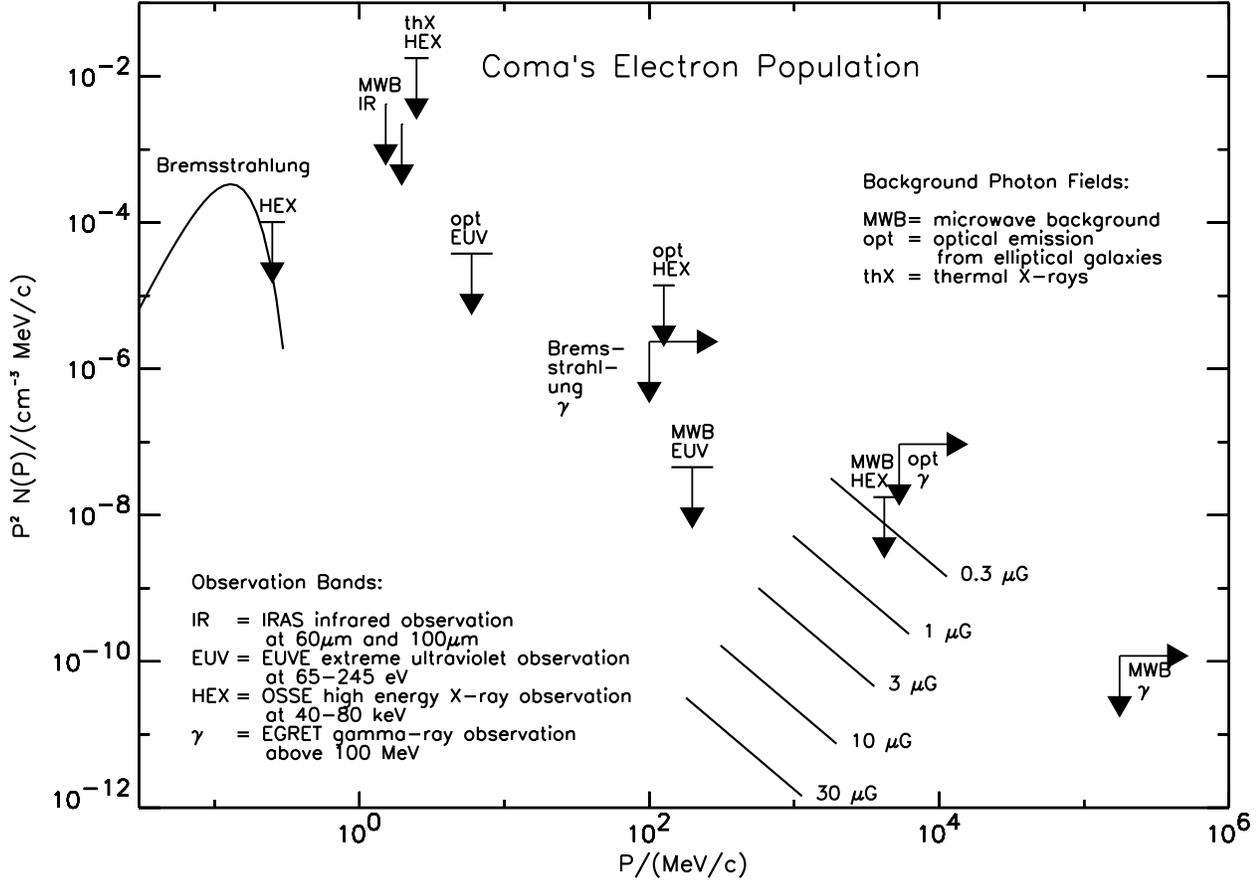,width=\textwidth,angle=0}
\caption[]{\label{fig:e}The central electron population of Coma. The
line labeled with `Bremsstrahlung' shows the thermal population. Also
limits from bremsstrahlung in the OSSE and EGRET bands are shown. The
other lines result from radio observations and are labeled with the
corresponding central field strength. Limits from inverse Compton
scattering are labeled with the relevant photon population and the
resulting energy range. All limits are two sigma limits except the EUV
limits, which result from a detected flux, and might be read as data
points, if the EUV emission results from inverse Compton scattering
instead from a cool ICM component. 
A horizontal line in this diagram corresponds to a power law $N_{\rm
e}(P_{\rm e}) \sim P_{\rm e}^{-2}$.}
\end{figure*}

\begin{table*}[btp]
\begin{center}
\begin{tabular}{|l|ll|c|c|c|c|c|}
\hline
	&&	& IR$_{60 \mu\rm m}$ & IR$_{100 \mu\rm m}$ & EUV$_{\rm
65-245eV}$ & HEX$_{40-80 \rm keV}$ & gamma$_{\rm >100 MeV}$\\
\hline
MWB	& $P_{\rm e}$ &$ [MeV/c]$         &1.5              & 2.0
        & $142 - 275$      & $(3.5-5.0)\cdot 10^3$ & $>\,1.8\cdot 10^5$ \\
	& $C_{\rm eo}$&$ [{\rm cm^{-3}\,h_{50}}]$ &$5.1\cdot 10^{-2}$
& $4.1\cdot 10^{-2}$& $2.1\cdot 10^{-3} $& 0.13 		  & 0.47 \\
\hline
optical photons & $P_{\rm e}$ &$ [MeV/c]$&  &
& $4.3 - 8.3$ &   $106-150     $ &$>\,5.3\cdot 10^{3}$\\
& $C_{\rm eo}$&$ [{\rm cm^{-3}\,h_{50}}]$ & &
& $4.4\cdot 10^{-3} $& $0.28 $& 1.0\\ 
\hline 
thermal X-rays & $P_{\rm e}$ &$ [MeV/c]$& & & & $ 2.0-3.0  $ & Klein-Nishina\\
& $C_{\rm eo}$&$ [{\rm cm^{-3}\,h_{50}}]$ & & & & $ 0.49 $& suppressed\\ 
\hline
\end{tabular}
\end{center}
\caption[]{\label{tab:limits} The IC limits on the normalization constant
$C_{\rm eo}$ of the power law electron distribution (Eq. \ref{eq:pow})
resulting from the different combinations of background photon field
and observation bands, and the typical electron momenta necessary for
this scattering.}
\end{table*}

\section{Inverse Compton Emission}
\subsection{Coma's Electron Population}

Only two regions of the electron spectrum are visible: the X-ray
emitting thermal bulk with a temperature of 8.2 keV and relativistic
electrons around 1 GeV visible in the radio. Outside of these regions
information can be obtained from the detection or nondetection of IC
flux resulting from electrons and background photon fields, mainly the
microwave background (MWB) and the starlight of the cluster members. The
radio photons of the halo do not contribute significantly to the IC
flux: The energy density of the radio emission is of the order of
$L_{\rm radio}/(4\pi\,c\,r_{\rm c}^2) \approx 10^{-7}\,{\rm
eV\,cm^{-3}}$ and therefore 6 orders of magnitude lower than that of
the MWB. The limits resulting from the different
combination of photon fields and observation bands are given
in Tab. \ref{tab:limits} and plotted in Fig. \ref{fig:e} above the
typical electron momenta necessary to scatter the peaks of the
blackbody spectra into the observation bands from the relation
$<\varepsilon > =  \frac{4}{3}\,\gamma_{\rm e}^2\, 2.70\,kT$
(Blumenthal \& Gould 1970)\nocite{blumenthal70}.  For the
calculations, which are explained below, the spectral index of the
radio electrons is used.

\subsection{Microwave Background\label{Sec:MWB}}
The expected IC flux from the MWB is (derived from Eq. (7.31) in Rybicki
\& Lightman (1979)\nocite{rybicki79} for Thomson scattering)
\begin{eqnarray}
\label{IC}
&& F_\gamma(>E_\gamma) = \frac{\tilde{V}_{\rm IC}}{4 \pi D^2}\,
\frac{8 \pi^2 r_{\rm e}^2}{h^3 c^2} (kT)^3 f_{\rm IC}(\alpha) C_{\rm eo}
\left(\frac{E_\gamma}{kT}\right)^{-\frac{\alpha-1}{2}}
\\ 
&& f_{\rm IC}(\alpha) =
\frac{2^{\alpha +4}(\alpha^2+4\alpha+11)}{(\alpha+3)^2(\alpha+5)(\alpha^2-1)} 
\,\Gamma{\textstyle\left(\frac{\alpha+5}{2}\right)}
\,\zeta{\textstyle\left(\frac{\alpha+5}{2}\right)}\\
&& 
\label{eq:VIC}
\tilde{V}_{\rm IC} = 2 \,\pi\, r_{\rm c}^3\,
{\cal B}_{\scriptscriptstyle \frac{R_{\rm o}^2}{R_{\rm o}^2+r_{\rm c}^2}}
{\textstyle \left( \frac{3}{2}, \frac{3}{2}
\left(\beta -1 \right) \right)}\,\, ,
\end{eqnarray}
where $\zeta$ denotes the Riemann zeta function, and $R_{\rm o}$ the
radius covered by the observation. The electrons seen in the radio
scatter microwave photons into the hard X-ray band. Thus, with the
help of Eq. \ref{eq:Ceo} we can use their limit in order to get a
lower limit on the central field strength of $B_{\rm o} > 0.2
\,\mu{\rm G}\, \phi_{B}^{-0.43} h_{50}^0$, which is independent of the
Hubble constant. Assuming a uniform magnetic field strength over the
whole cluster volume by setting $\beta = 0$, gives only $B_{\rm o} >
0.1 \,\mu{\rm G}\, \phi_{B}^{-0.43} h_{50}^0$, in agreement with the
value of Rephaeli et al. (1994)\nocite{rephaeli94}.  But this is an
unrealistic configuration, as explained in Sect. \ref{Sec:Radio}.

Since none of the halo formation theories discussed in
Sect. \ref{Sec:halo} predicts any break in the electron spectrum
between 100 MeV/c and GeV/c, it is reasonable to extrapolate the
spectrum and use the limit given by the EUV flux to predict a field
strength stronger than
\begin{equation}
B_{\rm o} > 1.2 \,\mu{\rm G} \left(  
\frac{\phi_{B}\,\, F_{EUV,IC-limit}}{0.16/( {\rm
cm^{2}\, s\, keV})} \right)^{-0.43}\,\,.
\end{equation}
If the EUV IC contribution can be further constrained, a higher field
strength would follow. A break by 0.5 in the radio spectral index, placed at
the lowest observed frequency of 30.9 MHz (Henning
1989)\nocite{henning89} still gives a limit $B_{\rm o}> 0.3 \,\mu{\rm
G}\, \phi_{B}^{-0.54}$.

\subsection{Optical Photons}
The luminosity of elliptical galaxies within the central $700 \,{\rm
arcmin^2}$ is given by an integration of the R-band luminosity
function of Secker \& Harris (1996)\nocite{secker96} plus the
luminosities of NGC 4874 and NGC 4889 (Strom \& Strom
1978)\nocite{strom78}, which are not included in this luminosity
function.  We assume that this radiation has a blackbody spectrum with
a typical temperature of 3000 K, and therefore use a bolometric
correction of $m_{\rm R} - m_{\rm bol} = 1.3$ (Webbink \& Jeffers
1969). The radial emission profile is that of the galaxy
distribution. $\varepsilon(r') \sim (1+(r'/r_{\rm G})^2)^{-
\alpha_{\rm G}}$, with $\alpha_{\rm G}=0.8$ and $r_{\rm G} = 160 \,{\rm
kpc}\, h_{50}^{-1}$ (Girardi et al. 1995), which we use up to a radius
of $R_{\rm G}= 5 \, {\rm Mpc}\, h_{50}^{-1}$. We estimate a central
bolometric emissivity of $\varepsilon_{\rm o} = 1.0\cdot 10^{13}\,
L_{\odot}\, {\rm Mpc^{-3}}\, h_{50}$.
In order to get the photon density field at a position $\vec{r}$, this
has to be folded with the radiation field of a source at $\vec{r}'$:
$n_{\gamma}(\vec{r},\vec{r}') =
\varepsilon(r')/(4\pi c|\vec{r}-\vec{r}'|^2)$. Integrating over
all possible angles between $\vec{r}$ and $\vec{r}'$ gives the photon
density
\begin{equation}
\label{eq:phF}
n_{\gamma}(r) = \frac{1}{2\,c\,r} \int\limits_o^{R_{\rm G}} dr'\, r'\,
\varepsilon(r') \, \ln \frac{r+r'}{|r-r'|}\,\,.
\end{equation}
Normalizing this by dividing with the photon density of a blackbody
cavity, and integrating it together with the normalized
$\beta$-profile of the relativistic electrons numerically over the
volume gives the effective volume $\tilde{V}_{\rm IC}$, which replaces
Eq. \ref{eq:VIC}. The scattering of optical photons into the EGRET
band needs electrons, visible in the radio. Thus a limit of $B_{\rm
o}> 0.1 \,\mu{\rm G}\, \phi_{B}^{-0.43}$ can be set.  The assumed
temperature of the radiation mainly enters this calculation by the
bolometric correction. The anisotropy of the optical photon field at a
given position does not change the resulting IC flux of the cluster,
since the integration over shells of constant radius averages the
different contributions.

\subsection{Thermal and Nonthermal X-Ray Photons}

The radial density profile of X-ray photons is also described by
Eq. \ref{eq:phF}, with an emissivity profile $\varepsilon(r') \sim
n_{\rm e}^2(r') \sim (1+(r'/r_{\rm c}))^{-3\beta}$. We calculated the
spectrum of the Bremsstrahlung using the Gaunt factor given in Novikov
\& Thorne (1972)\nocite{novikov72}, a central gas density of
$n_{\rm e,o} = 3\cdot 10^{-3}\,{\rm cm^{-3}\,h_{50}^{1/2}}$, and a gas
temperature of $kT =8.2$ keV (Briel et al. 1992\nocite{briel92}). The
IC scattering of this photons had to be calculated numerically. The
contribution to the EGRET band is negligible, due to the reduction of
the IC cross section in the Klein-Nishina regime. The calculation of
the scattering into the OSSE band was done using the exact probability
distribution of frequency shifts of Thomson scattering, valid from
non- to ultrarelativistic electron momenta, which is derived in the
Appendix.

The limit resulting from the expected bremsstrahlung in the OSSE band
of a suprathermal electron distribution is $C_{\rm eo} < 6 \cdot
10^{-5}\, {\rm cm^{-3}}\,h_{50}^{1/2}$. Bremsstrahlung of relativistic
electrons was calculated with the formulae in Blumenthal \& Gould
(1970)\nocite{blumenthal70}.  The EGRET limit translates into $C_{\rm
eo} < 3.3\cdot 10^{-2}\,{\rm cm^{-3}}\,h_{50}^{1/2}$. These limits are
also shown in Fig
\ref{fig:e}.  

\section{Radio Halo Formation\label{Sec:halo}}

The slope of the electron spectrum is determined by the acceleration,
cooling and injection processes. Three sources for the energetic
electrons have been discussed in the literature:
\begin{itemize}
\item
The {\it primary electron model} (Jaffe 1977, Rephaeli 1979), in which
the electrons diffuse out of radio galaxies and form the radio
halo. Their spectrum should be a straight power law. A steepening of
the spectral index appears at a break frequency, if there is some
momentum dependent escape from the halo region.
\item
In the {\it in-situ acceleration model} (Jaffe 1977, Roland et
al. 1981, Schlickeiser et al. 1987, Burns et al. 1994, Deiss et
al. 1997) the electrons are accelerated within the ICM by plasma waves
or shocks. Burns et al. (1994) report evidence for a recent merger
event of Coma $\sim 2$ Gyr ago, which was strong enough to power the
radio halo.  An exponential cutoff in the spectrum should occur at the
momentum, where the cooling time scale gets comparable to the
acceleration time scale.
\item
The {\it secondary electron model} (Dennison 1980) assumes the
relativistic electrons, but also positrons, to result from pion decay
after hadronic interaction of a relativistic proton population with
the background gas. Since the protons should result from radio
galaxies (En{\ss}lin et al. 1997)\nocite{ensslin97} and supernova
remnants in starburst galaxies (V\"{o}lk et al. 1996)\nocite{voelk96},
which are able to accelerate to high energies, no cutoff in the
electron spectrum is expected. But the injection spectrum peaks at
35 MeV/c (due to the kinematics of the decay $\pi^\pm \rightarrow
\nu_{\mu}\mu^\pm, \mu^\pm \rightarrow \nu_{\mu} \nu_{e}  e^\pm$), and
therefore the equilibrium spectrum should flatten to a spectral index
$\alpha = 2$ (due to synchrotron and IC cooling) below the main
injection energy range.
\end{itemize}
                           
Schlickeiser et al. (1987) report a high frequency cutoff in the radio
spectrum of the halo at 2.7 GHz, which would indicate a cutoff in the
electron distribution, and therefore supports the {\it in-situ
acceleration model}. But Deiss et al. (1997) claim that their much
higher measured flux at 1.4 GHz is not consistent with the value of
Schlickeiser et al. (1987). Deiss et al.  (1997) suspect that the
value of the flux given by these authors is much too low, which
implies that the claimed strong steepening of the spectrum above 1.4
GHz is not real. In the literature (e.g. Schlickeiser et al. (1987) or
Deiss et al. (1997)) citations of an upper flux limit at 5 GHz appear
frequently, measured by Waldthausen (1980)\nocite{waldthausen80}. But
his observation was restricted to the central region of the Coma
cluster, and therefore no proper background determination was
possible. In order to clarify the question about the existence of a
cutoff in the electron distribution, which is important for models of
halo formation, new high frequency observations of the diffuse radio
halo are necessary.

\begin{figure}
\psfig{figure=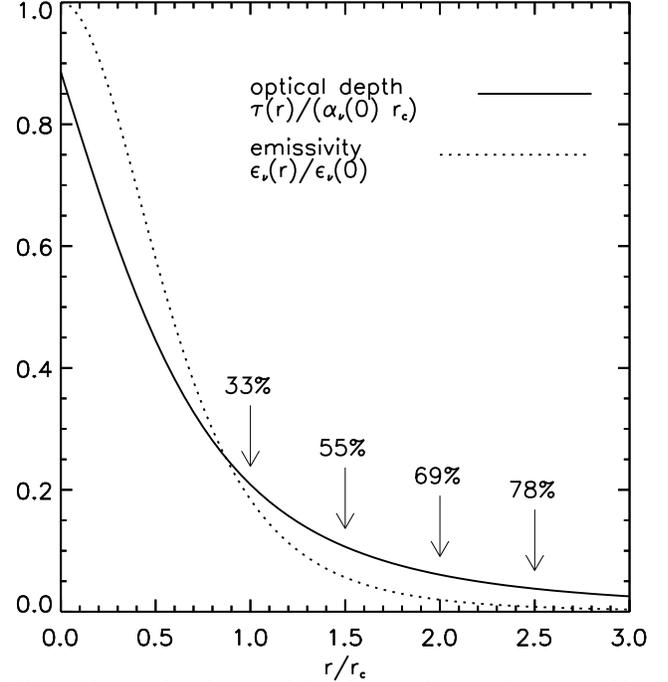,width=\textwidth,angle=0}
\caption[]{\label{fig:OptDept}Normalized optical depth $\tau(r)$ for
synchrotron self absorption (solid) and synchrotron emissivity
$\varepsilon_{\nu}(r)$ (dotted) as a function of radius. The numbers
indicate the percentage of total emission produced within the
indicated radius. A radial unlimited emission region was assumed for
simplicity.}
\end{figure}

The low energy spectral break expected in the {\it secondary electron
model} cannot be used easily for a distinction between the models,
since we expect a sharp step in the spectrum in the same energy range,
which could make it difficult to detect the appearance of the break.
This step is caused by a sharp change in the cooling time scale as a
function of energy, due to the transition from optical thin
synchrotron emission at higher energies to optically thick emission
below the step energy. In the optical thick regime, the radio power
emitted by the electrons is reabsorbed, and therefore synchrotron
cooling is suppressed. The synchrotron self absorption coefficient
(Eq. (6.53) in Rybicki \& Lightman 1979\nocite{rybicki79}) averaged
over an isotropic distribution of electron pitch angles is
\begin{eqnarray}
\alpha_\nu &=& \frac{\sqrt{3}\, e^3\, B\,C_{\rm e}}{8 \pi m_{\rm e}\,
\nu_{\rm o}^2} f_{\rm a}(\alpha)
\left( \frac{\nu}{\nu_{\rm o}}   \right)^{-\frac{\alpha+4}{2}}\\
f_{\rm a}(\alpha) &=& \frac{\sqrt{\pi}\,\Gamma(\frac{\alpha +6}{4})
\,\Gamma(\frac{3\alpha +2}{12}) \,\Gamma(\frac{3\alpha +22}{12})}{
2 \,\Gamma(\frac{\alpha +8}{4})}\,\,,
\end{eqnarray}
and can be evaluated at the cluster center using Eq. \ref{eq:Ceo}:
\begin{eqnarray}
\alpha_\nu(0) &=& \frac{342}{r_{\rm c}}\,
\left(\frac{\nu}{\rm 0.1\,MHz}  \right)^{-3.84} \left( \frac{B_{\rm
o}}{\mu{\rm G}} \right)^{\frac{1}{2}}\\
&=& \frac{3.9}{r_{\rm c}}\,
\left(\frac{P_{\rm e}}{\rm 0.1 \,GeV/c}  \right)^{-7.68} \left(
\frac{B_{\rm o}}{\mu{\rm G}} \right)^{-3.34}\,\,,
\end{eqnarray}
where the frequency was translated to electron momentum using the
monochromatic approximation $\nu = (\beta_{\rm e} \gamma_{\rm
e})^2\,\nu_o$. The optical depth $\tau(r)$ at a radius $r$ is given by
an integral from that radius to infinity over the radial profile of
the absorption coefficient
\begin{equation}
\alpha_{\nu}(r) \sim C_{\rm e}(r)\,B(r)^{\frac{\alpha +2}{2}} \sim
\left[ 1+ ( r/r_{\rm c})^2 \right]^{-\frac{3(\alpha
+6)}{8} \beta}\,\,,
\end{equation}
and is plotted in Fig. \ref{fig:OptDept}. Most of the total power at a
given frequency is emitted within 2 core radii, and would result from
a region with optical depth bigger than one if the central absorption
coefficient is $\alpha_\nu(0)> 1/(0.05 \,r_{\rm c})$. This happens for
$\nu < \tilde{\nu} = 21 \,{\rm kHz}\,\,(B_{\rm o}/{\mu{\rm G}})^{0.13}$
corresponding to $P_{\rm e} < \tilde{P}_{\rm e} = 81\, {\rm MeV/c}\,\,
(B_{\rm o}/\mu{\rm G})^{-0.43}$. This sudden change in the radio
spectrum can hardly be detected, but should make a signature in a
halo electron population, which was injected at higher energies and
cools by IC and synchrotron losses, as it is the case for the {\it
primary} and {\it secondary electron model}. The time independent
solution of the continuity equation for this electrons in momentum
space is
\begin{equation}
N_{\rm e}(P_{\rm e}) = \frac{1}{b(P_{\rm e})} \int\limits_{P_{\rm
e}}^\infty \,dp\,q_{\rm e}(p)\,\,,
\end{equation}
where $q_{\rm e}(p)$ is the injection spectrum and 
\begin{equation}
b(P_{\rm e}) = \frac{4}{3} \,\sigma_{\rm T} \,\left( \frac{P_{\rm e}
}{m_{\rm e} \,c} \right)^2 \, \left( \frac{B_{\rm o}^2}{8\,\pi}\,{\rm
H}(P_{\rm e}-\tilde{P_{\rm e}}) + U_{\rm ph} \right)
\end{equation}
the cooling function. $U_{\rm ph}$ is the photon energy density, which
is dominated by the MWB, and ${\rm H}$ is the
Heaviside step function, used as an approximation of the sharp change
in opacity. The height of the step is therefore
\begin{equation}
\lim_{\delta\to 0}\, \frac{N_{\rm e}(\tilde{P}_{\rm e} -
\delta)}{N_{\rm e}(\tilde{P}_{\rm e} + \delta)} =
\frac{B_{\rm o}^2}{8\pi\, U_{\rm ph}}+1
\end{equation}
The exact shape of the transition is a little bit more complicated to
calculate, since the problem is nonlinear due to the mutual dependence
of the electron spectrum and the absorption coefficient, the geometry
of the radiation transport, which has to be considered, and the time
dependence before establishing the equilibrium spectrum.  For a field
of e.g. $B_{\rm o} = 10 \,\mu$G a step in the electron spectrum of one
order of magnitude is expected at $P_{\rm e} = 26
\,$MeV/c (using our simple approach), very close to the position of
the expected flattening below the injection peak in the {\it secondary
electron model}. The {\it in-situ acceleration model} predicts no step
at this energy, since the cooling function determines only the high
energy cutoff region of the spectrum, where the cooling gets
comparable to the acceleration. We note, that any break in the
electron spectrum below the observed range would lower the effect of
synchrotron self absorption. Assuming a break in the radio spectrum at
30.9 MHz as in Sect. \ref{Sec:MWB} shifts the step in the above example
to 15 MeV/c.

\section{Conclusion}

The search for diffuse emission in any photon energy band leads to
interesting information about the shape of the relativistic electron
population in the Coma cluster of galaxies. We present a compilation
of several limits, and discuss spectral features of the electrons
within the various radio halo formation models. A signature in the low
energy range is expected to arise from the sharp change in opacity for
low frequency radio photons due to synchrotron self absorption, which
could distinguish between a relativistic electron population which is
cooling and one which is gaining energy, as it is predicted for {\it
primary} and {\it secondary electron models} on the one hand, and for
{\it in-situ acceleration models} on the other hand. Another
signature of the {\it secondary electron model} is a flattening of the
spectrum below the main electron injection energy. Future observations
of diffuse emission might test the electron distribution sensitively
enough in order to see these signatures. Present observations restrict
the electron population sufficiently, so that a central magnetic field
strength higher than $1.2 \,\mu{\rm G}$ is needed in order to explain
the observed synchrotron emission, if the steep volume integrated
radio spectrum with spectral index of 1.34 can be extrapolated to
lower frequencies, or a limit of $B_{\rm o}> 0.3
\,\mu{\rm G}$, if a break  appears below the observed range.
Since these limits arise from a detected EUV flux excess, which is
interpreted to result from a cool gas component of the intra cluster
medium, the possibility remains, that this flux results in fact from
inverse Compton scattering of microwave photons, if the field strength
is below $1.2\,\mu$G.  This possibility seems to be in contradiction
with the Faraday measurement of Feretti et al. (1995) of $6\,
\mu$G. But since the inverse Compton argumentation gives volume
averaged field strength which are weighted with the relativistic
electron distribution, stronger field strength in regions with a low
density of relativistic electron could solve this contradiction. If
the nature of this EUV excess can be understood theoretically, the EUV
limit on inverse Compton flux might be improved and a higher field
strength limits might result.

Limits without extrapolation of the electron spectrum come from the
scattering of starlight photons to EGRET energies ($B_{\rm o}> 0.1
\,\mu{\rm G}$) and microwave photons to OSSE energies ($B_{\rm o}> 0.2
\,\mu{\rm G}$). Magnetic field limits from inverse Compton limits are
below the estimates resulting from Faraday rotation measurements,
being therefore in good agreement. They measure the magnetic content
of the Coma cluster on a more global scale, and demonstrate
independently that the observed Faraday rotation results from magnetic
fields within the intra cluster medium. Future EUV, X- and gamma-ray
telescopes will raise these limits on the magnetic field strength in
the Coma halo region to the level of Faraday measurements, or they
will detect energetic nonthermal photons from Coma, resulting from
inverse Compton scattering or from other sources.\\[0.5em]

\noindent
{\it Acknowledgments.} T.A.E. is supported by the {\it Studienstiftung
d. dt. Volkes}. We acknowledge discussions with Richard Lieu and
useful comments by Jack O. Burns.

\begin{appendix}
\section{Transrelativistic Thomson Scattering}

We follow the derivation of the formulae of transrelativistic Thomson
scattering by Wright (1979).  Scattering of a photon from an isotropic
photon field by an electron with velocity $\beta_{\rm e}\,c$ and
energy $\gamma_{\rm e}\,m_{\rm e}\,c^2$ has a probability distribution
of the angle $\theta$ between photon and electron direction in the
electron's rest frame given by
\begin{equation}
f(\mu)\,d\mu = [2 \gamma_{\rm e}^4(1-\beta_{\rm e}\mu)^3]^{-1}\,d\mu\,\,,
\end{equation}
where $\mu =\cos\theta$. The logarithmic frequency shift from $\nu$ to
$\nu '$ of the photon after scattering into an angle $\theta '$  is
\begin{equation}
\label{eq:s}
s = \ln\left( \frac{\nu '}{\nu} \right) = \ln\left(\frac{1+\beta_{\rm e}\mu
'}{1-\beta_{\rm e}\mu} \right)\,\,.
\end{equation}
The probability distribution of $\mu ' = \cos \theta '$ for a given
$\mu$ is 
\begin{equation}
g(\mu '; \mu)\,d\mu' = \frac{3}{8} [1+\mu^2 \mu'^2 +\frac{1}{2}
(1-\mu^2)(1-\mu'^2)] \,d\mu'\,\,.
\end{equation}
The probability for a shift $s$ given the electron velocity is
\begin{equation}
\label{eq:Pint}
P(s;\beta_{\rm e})\,ds = \left[ \int d\mu \,f(\mu)\,g(\mu ';\mu)\,
\left( \frac{\partial \mu '}{\partial s} \right) \,\right]\,ds \,\,.
\end{equation}
$\mu'(s,\beta_{\rm e},\mu)$ is given by (\ref{eq:s}) and the range of
integration is given by the conditions $-1<\mu<1$ and $-1<\mu'<1$.
Rephaeli (1995) simplified this integral by assuming that the
scattering is isotropic in the electron's rest frame, which allows to
average over the angle between background and scattered photon. He
compared scattered photon spectra using the simplified and exact
formulae numerically, and found no significant difference for electrons
with energies of 1...15 keV, typical for the thermal component of the
ICM. But suprathermal electrons, expected to be the connection between
the thermal and the halo electrons, see an anisotropic photon field in
their rest frame.  Fortunately, the integrand in (\ref{eq:Pint}) is a
rational function of $\mu$ and therefore analytically
integrable. Integrating the two regimes $s<0$ and $s>0$ separately and
introducing the dimensionless momentum $p=\beta_{\rm e}\,\gamma_{\rm
e}$ leads to:
\begin{eqnarray}
{\textstyle P(s,p)} &{\textstyle\, =\,}&
\frac{3 e^s (1+e^s)}{8 p^5 } \left[
\frac{3+3p^2+p^4}{\sqrt{1+p^2}} - \frac{3+2p^2}{2p} {\textstyle
(\tilde{s} - |s|)} \right]   \nonumber 
\end{eqnarray}
\begin{eqnarray}
\label{eq:myP}
&&+
\frac{3 {\rm sgn}(s) }{32p^6} \left[ {\textstyle 1-e^{3s} + 3 e^s (1-e^s)
(9+8p^2+4p^4)} \right] 
\end{eqnarray}
The maximal logarithmic frequency shift is given by
\begin{equation}
\tilde{s} = \ln \left( \frac{1+\beta_{\rm e}}{1-\beta_{\rm e}}
\right)= \ln \left( \frac{\sqrt{1+p^2}+p}{\sqrt{1+p^2}-p} \right)\,\,,
\end{equation}
and therefore is $P(s,p)=0$ for $|s|>\tilde{s}$.
The distribution of frequency shifts of scattering with a spectrum of 
electrons $n_{\rm e}(p)\,dp$ 
\begin{equation}
P(s)\,ds = \frac{\int\, dp \,n_{\rm e}(p)\,P(s;p)}{\int\, dp \,n_{\rm
e}(p)} \,ds
\end{equation}
has to be folded with the photon spectrum $n(\nu)\,d\nu$ in order to
obtain the source spectrum of scattered photons 
\begin{equation}
q(\nu')\,d\nu' = \sigma_{\rm T}\,c\,n_{\rm e}\, \left[
\int_{-\infty}^{+\infty} \, ds \, P(s)\,
n(\nu'\,e^{-s}) \,\right] d\nu'\,\,.
\end{equation}
These formulae can be applied from non- to ultrarelativistic electron
momentum, they are valid for scattering to higher and lower energies
compared with the background photon energy, but they are restricted to the
Thomson regime ($h\,\nu \ll \gamma_{\rm e} \,m_{\rm e}\,c^2$). Applications
are especially IC scattering by transrelativistic electrons, such as
suprathermal electrons, or electrons of a very hot plasma. Since
the evaluation of (\ref{eq:myP}) needs only a little more computation time
than its approximations, it could be included in any numerical
code for Thomson scattering.

\end{appendix}

 
\end{document}